\documentclass[aps,prd,epsf,showpacs,amsmath,amssymb,graphics,twocolumn,preprint,10pt]{revtex4}
\usepackage{graphicx}
\usepackage{dcolumn}
\usepackage{bm}

\newcommand{\be}{\begin{equation}}
\newcommand{\ee}{\end{equation}}
\newcommand{\ba}{\begin{eqnarray}}
\newcommand{\ea}{\end{eqnarray}}
\newcommand{\ban}{\begin{eqnarray*}}
\newcommand{\ean}{\end{eqnarray*}}

\newcommand{\eq}[1]{(\ref{#1})}

\begin{document}
\title{Acceleration of particles and shells by Reissner-Nordstr\"{o}m naked singularities}
\author{$^1$Mandar Patil\footnote{Electronic address: mandarp@tifr.res.in},
$^1$Pankaj S. Joshi\footnote{Electronic address: psj@tifr.res.in},
$^2$Masashi Kimura\footnote{Electronic address: mkimura@yukawa.kyoto-u.ac.jp}, 
$^3$Ken-ichi Nakao\footnote{Electronic address: knakao@sci.osaka-cu.ac.jp}
}

\affiliation{$^1$Tata Institute of Fundamental Research, Homi Bhabha Road, Mumbai 400005, India.\\
$^2$Yukawa Institute for Theoretical Physics, Kyoto University, Kyoto 606-8502, Japan. \\
$^3$Department of Mathematics and Physics, Graduate School of Science,
  Osaka City University, Osaka 558-8585, Japan. 
}


\begin{abstract}
We explore the Reissner-Nordstr\"{o}m naked singularities with a charge $Q$ larger than 
its mass $M$ from the perspective of
the particle acceleration. We first consider a collision between two test particles
following the radial geodesics in the Reissner-Nordstr\"{o}m naked singular geometry. 
An initially radially ingoing particle turns back due to the repulsive effect
of gravity in the vicinity of naked singularity. Such a particle then collides with an another radially
ingoing particle. We show that the center of mass energy of collision
taking place at $r \approx M$ is unbound, in the limit where
the charge transcends the mass by arbitrarily small amount $0<1-M/Q\ll1$.
The acceleration process we described avoids fine tuning of the parameters of the particle geodesics for 
the unbound center of mass energy of collisions and the proper time required for the process is also finite.
We show that the coordinate time as observed by the distant observer required for the trans-Plankian 
collisions to occur around the naked singularity {\sf with one solar
mass} is merely of the order of million years
which is much smaller than the Hubble time. On the contrary, the 
time scale for collisions 
associated with extremal black hole in an analogous situation
is many orders of magnitude larger than the age of the universe.  
We then study the collision of the neutral spherically symmetric shells made up of
dust particles. In this case, it is possible to treat the situation by exactly taking into account 
the gravity due to the shells using Israel`s thin shell formalism, and thus this treatment allows us to go 
beyond the test particle approximation. The center of mass energy of collision of the shells 
is then calculated in a situation analogous to the test particle case
and is shown to be bounded above. However, we find that
the energy of a collision between two of constituent particles of the
shells at the center of mass frame can exceed
the Planck energy.

\end{abstract}
\preprint{OCU-PHYS-355}
\preprint{AP-GR-93}
\preprint{YITP-11-69}


\maketitle

\section{Introduction}
Since the terrestrial particles accelerators like Large Hadron
 Collider probe particle physics
at the energy scales that are almost $15$ orders of magnitude smaller than the Planck scale, 
it would interesting to investigate whether or not various naturally occurring high energy astrophysical
phenomenon could shed light on the new physics at higher energy scales that remain unexplored. 
Stepping ahead towards this exciting
possibility,  an interesting proposal was made recently which suggests that the Kerr black holes 
could act as particle accelerators\cite{BSW}. It was shown that the two particles dropped in from infinity
 at rest, traveling along the timelike geodesics can collide and interact near the event horizon of a 
Kerr black hole with divergent center of mass energy, provided the black hole is close to being extremal 
and angular momentum of one of the particles takes a specific value of the orbital angular momentum. 
The possible astrophysical implications of this process 
around the event horizon of the central supermassive black hole 
in the context of annihilations of the dark matter particles accreted from the galactic halo 
were also investigated \cite{BSW2}. This process of particle acceleration suffers from several 
drawbacks and limitations pointed out in\cite{Berti}. The angular momentum of one of the colliding 
particle must take a single fine tuned value. The proper time required for the particle with 
fine tuned angular momentum to reach the horizon and thus the time required for the collision to 
take place is infinite. The gravity produced by the colliding particles themselves was neglected.
There were many investigations of this acceleration mechanism in the background of Kerr as well 
as many other black holes\cite{BHaccel}.

Two of present authors, PM and PSJ, 
studied and extended the particle acceleration mechanism to the 
Kerr naked singular geometries transcending
Kerr bound by arbitrarily small amount
$0<a-1\ll1$ \cite{Patil}. We considered two different scenarios where the colliding particles follow 
a geodesic motion along the equatorial plane as well as along the axis of symmetry of the Kerr geometry.
In the first case, the particles are released from infinity at rest in the equatorial plane.
One of the initially infalling particle turns back as an outgoing particle due to its angular momentum. 
It then collides with an another infalling particle around $r=1$. We showed that the center of mass 
energy of collision between these two particles is arbitrarily large. The angular momentum of the 
colliding particles is required to be in a finite range as opposed to the single fine tuned value
in case of Kerr black holes. Thus the extreme fine tuning of the
angular momentum is avoided in such a collision.

 The proper time required for such a collision to take place is also shown to be finite. In the second case, 
the particles are released from rest along the axis of symmetry, from large but finite distance. 
These particles have zero angular momentum. One of the particles initially falls in and then turns 
back due to the repulsive effect of gravity in the vicinity of a Kerr naked singularity. This particle 
then collides with an ingoing particle at $z=1$. The center of mass energy of collision is arbitrarily 
large and the proper time required for the process to take place is finite. Thus two issues related to 
acceleration mechanism in Kerr black hole case, namely the fine tuning of the angular momentum and the 
infinite time required for the collision, are avoided in case of Kerr naked singularities.

The issue of the self-gravity of the point particles is difficult to deal in general. 
The accretion of the particles onto an astrophysical object can be expected to be more or less 
isotropic in many cases. Thus it would be interesting and more physical to study the motion and collisions 
of the shells of particles instead. The rigorous mathematical analysis of the shells would be very extremely 
difficult in the Kerr spacetime due to the lack of sufficient symmetry. 
By contrast, the motion and collision of the spherical shells would be exactly tractable in the spherically 
symmetric spacetimes following the Israel`s thin shell formalism\cite{Israel}. We first note that while no 
gravitational radiation is emitted by a perfectly spherical shell, the gravitational radiation per particle 
emitted by a quasispherical shell of particles will be significantly lower than the radiation emitted by 
a single particle\cite{Nakamura}. Thus it might be reasonable to ignore
the gravitational radiation effects and focus entirely on the backreaction while dealing with the shells.

The acceleration of the particles around the extremal Reissner-Nordstr\"{o}m black hole was studied in 
\cite{Zaslavskii},\cite{Nakao}. This process is mathematically similar
to the acceleration process in Kerr geometry.
The center of mass energy of collision near the horizon of the extremal Reissner-Nordstr\"{o}m black hole,
 of the charged and uncharged particles is shown to be divergent. The collision of the charged and uncharged 
spherical shells was investigated in\cite{Nakao}. The dynamics of the
shells when their gravity is ignored is
 same as that of the test particles. Whereas when the exact
 calculation is done taking into account the self-gravity effects,
 the center of mass energy turns out to be finite. Thus it was speculated that the center of mass energy of 
collision of particles around Kerr black hole might also turn out to be finite when the gravity due to the
 colliding particles is taken into account.

In this paper we first describe the particle acceleration process in the background of 
Reissner-Nordstr\"{o}m naked singularities. We show that the center of mass energy of collision 
between two uncharged particles, one of them initially ingoing and other one initially ingoing, 
but turning back due to the repulsive effect of gravity in the
vicinity of naked singularity is arbitrarily large, 
when the collision happens around $r\approx M$, provided that the
deviation of the 
Reissner-Nordstr\"{o}m charge from 
the mass is extremely small. We calculate the coordinate time as
  seen by the 
distant observer, associated with the ultra-high
energy collisions for extremal black hole as well as for naked
singularity. 
We show that the time scale associated with the 
trans-Plankian collisions around naked singularity {\sf with one solar
  mass} is 
of the order of million years which is significantly smaller than the 
Hubble scale, whereas the timescale for the extremal black hole 
{\sf with the same mass as that of the naked singularity} 
is fifteen orders of magnitude larger than the age of the universe. Thus 
collision process around black hole suffers from the inflating 
timescale problem while such issue is absent in case of the naked
singularity. We then investigate
the collision between two uncharged shells made up of dust particles,
in a situation analogous to the particle collision, taking into account their 
gravity. We find that the center of mass energy of a collision between the shells
is bounded above. However, the center of mass energy of a collision
between two of constituent particles of the shells can exceed
the Planck energy which might be a threshold value of the quantum gravity.

In this paper, we adopt the geometrized unit in which the speed of light and 
Newton's gravitational constant are unity. 

\section{Acceleration of particles by Reissner-Nordstr\"{o}m naked singularities}

\subsection{Geometry of Reissner-Nordstr\"{o}m spacetime}

The Reissner-Nordstr\"{o}m spacetime 
is a unique solution of Einstein equations under the assumptions 
of spherical symmetry, asymptotic flatness
with the $U$(1) gauge field as a source of spacetime curvature.
The line element of the Reissner-Nordstr\"{o}m geometry in the 
spherical polar coordinates is given by
\begin{eqnarray}
ds^2=-f (r)dt^2+\frac{1}{f(r)}dr^2+r^2\left(d\theta^2+\sin^2\theta d\phi^2\right).  \label{RN}
\end{eqnarray}
where
\begin{equation}
f(r)=1-\frac{2M}{r}+\frac{Q^2}{r^2}.
\end{equation}
The gauge field is given by 
\begin{equation}
A_\mu=\frac{Q}{r}\delta^t_\mu.
\end{equation}
This solution contains two parameters $M$ and $Q$, namely the mass and $U$(1) 
charge. In this paper, we assume that $M$ and $Q$ are positive, 
\begin{equation}
M>0~~~{\rm and}~~~Q>0.
\end{equation}

In the Reissner-Nordstr\"{o}m spacetime, there is a spacetime singularity at 
$r=0$. This singularity is timelike and thus is necessarily locally naked.  
The location of the horizon in the Reissner-Nordstr\"{o}m spacetime 
is given by a solution to the equation $f(r)=0$ . 
There are two roots to this quadratic equation given by
\begin{equation}
r=r_{\pm}:=M \pm \sqrt{M^2-Q^2}.
\end{equation} 
There are two real positive roots to the equation if $M>Q$. The larger root $r=r_+$ 
is the location of the event horizon and this spacetime corresponds to 
a spherically symmetric charged black hole. 
The smaller root $r=r_-$ corresponds to the Cauchy horizon associated with 
the timelike singularity at $r=0$.  
If $M=Q$, there is only one positive root. In this case the black hole is
known as the extremal black hole with a degenerate event horizon at $r=M=Q$. 
In the case of $M<Q$, there is no real root to the equation $f(r)=0$. 
Thus, the event horizon is absent and the timelike singularity at $r=0$ is exposed 
to the asymptotic observer at infinity. This configuration thus contains a globally 
visible naked singularity. We will investigate the last case in this paper from the 
perspective of particle acceleration. 

Before proceeding further, it is worthwhile to mention that, the naked singularities are associated with pathological features like the breakdown of predictability and so on. That was precisely the reason  why Penrose came up with the cosmic censorship conjecture abandoning the existence of naked singularities in our universe\cite{Penrose}. However there were recent developments in the framework in string theory, which suggests by means of the specific worked out examples, that the naked singularities might be resolved by high energy stringy modification to the classical general relativity \cite{Gimon} and various pathological features disappear. This renders the classical naked singular solutions legal as long as one stays sufficiently away from high curvature region where quantum gravity would prevail.

\subsection{Motion of a test particle}

We now study the motion of a point test particle following a timelike geodesic in the Reissner-Nordstr\"{o}m 
spacetime. Let $\bm{U}$ be the 4-velocity of the particle. 
Without loss of generality, we assume that the motion of the particle is confined to the 
equatorial plane $\theta=\pi/2$. All of the metric components \eq{RN} are manifestly 
independent of time coordinate and azimuthal angular coordinate. 
This means that both of the time coordinate basis $\partial/\partial t$ and 
azimuthal angular coordinate basis $\partial/\partial \phi$ are Killing vectors
The following quantities are conserved along the geodesic of the particle
\begin{equation}
E:=-\bm{U}\left(\frac{\partial}{\partial t}\right)~~{\rm and}~~
L:=\bm{U}\left(\frac{\partial}{\partial \phi}\right).
\end{equation}
$E$ can be interpreted as the conserved energy of the particle per unit mass and $L$ can be 
interpreted as the conserved angular momentum of the particle per unit mass. Using these
 constants of motion and the normalization condition for 4-velocity 
of the particle, the components of the 4-velocity $\bm{U}$ are written as
\begin{eqnarray}
\nonumber
U^{t}=\frac{E}{f}\\
\nonumber
U^{r}= \pm \sqrt{E^2-f \left(1+\frac{L^2}{r^2}\right)}\\
U^{\theta}=0\\
\nonumber
U^{\phi}=\frac{L}{r^2}
\end{eqnarray}
$\pm$ stands for the radially outgoing and infalling particles respectively.
The second one in the above equations can also be written in the following form
\begin{equation}
\left(\frac{dr}{d\tau}\right)^2+V_{\rm eff}=E^2, \label{p-EOM}
\end{equation}
where $\tau$ is the proper time of the particle, and 
\begin{equation}
V_{\rm eff}=f\left(1+\frac{L^2}{r^2}\right).
\end{equation}
$V_{\rm eff}$ can be thought of as a effective potential. 

For simplicity and from the perspective of the comparison to shell collision that
 would be discussed in the next section, we assume that the angular momentum of the 
particle is zero $L=0$. This implies that the motion of the particle is purely radial.
The effective potential now can be written as
\begin{equation}
V_{\rm eff}=f=1-\frac{2M}{r}+\frac{Q^2}{r^2}
\end{equation}

\begin{figure}
\begin{center}
\includegraphics[width=0.5\textwidth]{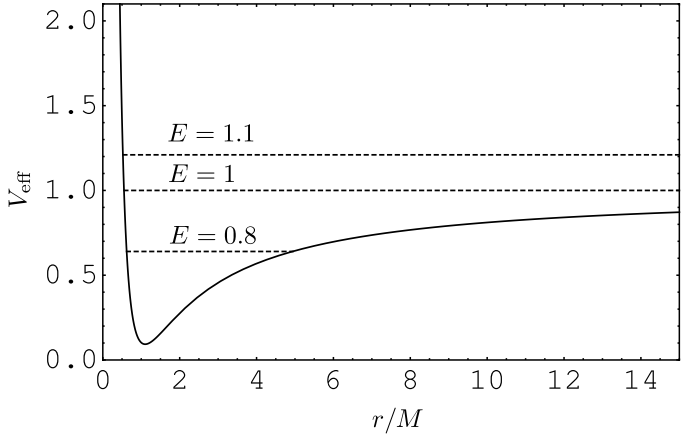}
\caption{\label{Veff_particle}
The effective potential is plotted as a function of $r/M$ for a test particle following a radial geodesic
in Reissner-Nordstr\"{o}m naked singular geometry with $Q/M=1.05$. 
An allowed domain for the motion of a particle is depicted by a dashed line for each case of specific energy.  
It admits a minimum at the classical radius $r=Q^2/M$, depicted by 'min', where gravity changes 
it's character from being attractive to repulsive in the close neighborhood of singularity. 
The ingoing particle thus gets reflected back as an outgoing particle close to singularity. The motion of a 
particle having energy $E=0.8<1$ is bound and oscillates.  
The motion of a particle with energy $E=1.1>1$ is unbound, has only one turning point. 
The motion of a particle with $E=1$ 
is marginally bound, also has only one turning point. The potential energy curve asymptotes to 
the $E=1$ as $r\rightarrow \infty$.
}
\end{center}
\end{figure}
The effective potential is plotted as a function of radius $r$ in Fig.~\ref{Veff_particle}.
For large values of radial coordinate 
$r\rightarrow \infty$, we have $V_{\rm eff}\rightarrow 1$. As one approaches the 
naked singularity $r\rightarrow 0$, effective potential blows up positively, 
i.e., $V_{\rm eff}\rightarrow \infty$. 
It always remains greater than zero and admits a minimum at $r=r_{\rm min}$ which is given by
\begin{equation}
r_{\rm min}=\frac{Q^2}{M},
\end{equation}
and we have
\begin{equation}
V_{\rm eff}|_{r=r_{\rm min}}=1-\frac{M^2}{Q^2}
\end{equation}
Note that $r_{\rm min}$ coincides with the classical radius associated with an object of 
charge $Q$ and mass $M$.
It is clear from the shape and slope of the effective potential curve that the gravity of the
Reissner-Nordstr\"{o}m
naked singularity is attractive in the domain $r_{\rm min}<r<\infty$, from the 
classical radius all the way upto infinity. Whereas the gravity is repulsive in 
the region extending from the singularity
to the classical radius $0<r<r_{\rm min}$.  
Similar behavior is also observed in case of other known examples of the 
stationary naked singularities\cite{Quevedo}.
An ingoing particle at initially speeds up upto the classical radius. 
It then slows down due to the repulsive gravity and gets reflected back eventually. 
It them emerges as an outgoing particle.

If the conserved energy of the particle is 
less than unity $E<1$ then the particle is bound, i.e., it oscillates back and forth
in the radial domain $b_-\leq  r \leq b_+$, where
\begin{equation}
b_\pm=\frac{M}{1-E^2}\left(1\pm\sqrt{1-\frac{Q^2}{M^2}\left(1-E^2\right)}\right).
\end{equation}
In the case of $E=\sqrt{1-M^2/Q^2}$, $b_+$ is equal to $b_-$. This means that 
the particle stays stably at rest at the classical radius $r=Q^2/M$. 
If the conserved energy is identical to unity $E=1$, then there is only one turning
 point given by $r=Q^2/2M$. In this case, the particle is at rest at infinity, and the motion of the 
particle is said to be marginally bound. In the case when energy is larger than unity 
$E>1$, again there is only one turning point given by $r=b_-$ since $b_+$ is negative in this case. 
The asymptotic velocity of the particle as it reaches infinity is 
positive $U^r\rightarrow \sqrt{E^2-1}$. Such a particle trajectory is called the unbound one. 

We should note that there is an important difference between the black hole case $M\geq Q$ 
and the naked singular case $M<Q$. 
In the case of the black hole $M\geq Q$, 
the radial motion cannot be restricted to only one asymptotically flat region. 
Since the inner turning point, $r=b_-$ is less than or equal to the radius of the Cauchy horizon 
$r=r_-$, the particle cannot return to the asymptotically flat region where it comes 
from. 
By contrast, in the case of the naked singularity $M>Q$, there is only one asymptotically flat region. 
Hereafter, we focus on the naked singular case $M>Q$. 

\subsection{Collision of test particles}

We now consider a collision between two particles moving along a 
radial geodesics i.e., $L=0$, each with mass $m$ and conserved energy $E=1$: Particles
 are assumed to be marginally bound, or in other words, they are released from rest from infinity.
 One could replace marginally bound particles by either unbound or bound particles. 
It does not change the conclusions. Let $U_1^\mu$ and $U_2^\mu$ be components of their 
4-velocities with respect to the coordinate basis.
We assume that one of the particles is initially ingoing particle which gradually
 slows down and eventually turns back as an outgoing particle due to the repulsive 
gravity in the vicinity of the naked singularity. Such a particle then collides with 
another ingoing particle at the 
radial coordinate $r$. By the assumption, $U_1^\mu$ and $U_2^\mu$ are given by
\begin{eqnarray}
\nonumber
U_1^\mu=\left(\frac{1}{f},\sqrt{1-f},0,0\right)\\
U_2^\mu=\left(\frac{1}{f},-\sqrt{1-f},0,0\right)
\end{eqnarray}
The energy of a collision between two particles at the center of mass frame is then given by \cite{BSW}
\begin{equation}
E_{\rm cm}^2=2m^2\left(1-g_{\mu\nu}U_1^\mu U_2^\nu\right)=\frac{4m^2}{f(r)}
=\frac{4m^2}{V_{\rm eff}},
\end{equation}
where $g_{\mu\nu}$ is the metric tensor given in Eq.~(\ref{RN}). 
It is seen from the above equation that the center of mass energy $E_{\rm cm}$ of collision depends on 
the location for the collision, for  given values of charge $Q$ and mass $M$. 
$E_{\rm cm}$ takes maximum when the effective potential $V_{\rm eff}$ takes minimum. 
The minimum of $V_{\rm eff}$ is realized at the classical radius $r_{\rm min}=Q^2/M$. 
If the collision takes place at $r=r_{\rm min}$, $E_{\rm cm}$ is given by
\begin{equation}
E_{\rm cm,max}^2=\frac{4m^2}{1-M^2/Q^2}
\label{ens}
\end{equation}
$E_{\rm cm,max}$ depends on the ratio of mass to the charge of Reissner-Nordstr\"{o}m spacetime. 
$E_{\rm cm,max}$ is very large if the charge transcends the mass by infinitesimally small amount.
 Here, we introduce a parameter defined by
\begin{equation}
\epsilon:=1-\frac{M}{Q}. \label{e-def}
\end{equation}
In the limit
$\epsilon\rightarrow 0$ ,
t$E_{\rm cm,max}$ becomes infinite,
\begin{equation}
\lim_{\epsilon \to 0}  E_{\rm cm,max}^2=\frac{2m^2}{\epsilon}\rightarrow \infty .
\end{equation}
 The above equation implies that the energy of collision measured at the center of mass frame 
 would be arbitrarily large.

In case of the black hole, the divergence of center of energy in the collision has
been demonstrated in near extremal or extremal geometries when the mass transcends the charge 
by arbitrarily small amount
$\epsilon\rightarrow 0^{-}$ .
In this paper, we have shown the possibility of the indefinitely large center of mass energy
in the naked singular geometry, which can be thought to be near extremal,
with the charge transcending the mass  by arbitrarily small amount
$\epsilon\rightarrow 0^{+}$ .

\subsection{Time scale of the collision}
We now estimate the time scale associated with the ultra-high energy particle collisions
in the Reissner-Nordstr\"{o}m naked singular geometry as well as in
the extremal black 
hole geometry and make
a critical comparison. We compute the proper time in the reference 
frame attached to the colliding neutral particle, as well as the
coordinate 
time measured by a distant static 
observer, required for the particle to reach the collision point 
$r=r_{\rm min}=Q^2/M$ in the case of naked singularity, and 
the horizon $r=M$ in the extremal black hole case. 
The particle starts from a distant location with $r_{\rm i}>r_{\rm min}$ and 
participates in the high energy collision.

In the extremal Reissner-Nordstr\"{o}m black hole geometry with $Q=M$, 
the high energy collision between the particles 
takes place at a location extremely close to the event horizon. 
One of the colliding particles is charged and the other one is charge 
neutral. The charged particle experiences an outward
repulsive electromagnetic force during its inward motion. 
For such a particle it turns out that 
$V_{\rm eff}=V^{'}_{\rm eff}=0$ as it approaches the event horizon,
as a consequence of which the proper time required for it to reach the horizon and 
participate in the high energy collision turns out to be infinite. The
neutral particle, 
however, falls 
freely following a geodesic motion and reaches the event horizon in 
a finite proper time as we show later in this section.
We also estimate the coordinate time as seen by the static observer at
infinity, 
required for the neutral particle to 
participate in the high energy collision. We show that it diverges 
in the limit of approach to 
the horizon as it is an infinite blueshift/redshift surface and the 
timescale associated with the trans-Plankian collision is 
much larger than the age of the universe.

In the Reissner-Nordstr\"{o}m naked singular geometry, the collision 
is between two charge neutral particles 
following a geodesic motion as they fall freely under the gravity. 
Both the conditions mentioned above in the last paragraph
 namely $V_{\rm eff}=V^{'}_{\rm eff}=0$ 
are not satisfied simultaneously anywhere along the trajectory of either of the two particles. Thus
the proper time required for the collision to take place for both the particles in their own frame 
is finite as we demonstrate later in this section. 
However, since it is necessary to have $f(\rm r_{min}) \rightarrow 0^{+}$, for high energy 
collision to occur, which is precisely the condition for extremely large blueshift/redshift,
one would expect that coordinate time as measured by the static observer at infinity would diverge. 
We show that for trans-Plankian collisions the coordiante time
required 
is of the order of million years which is 
much smaller than the Hubble time.

For a particle moving along a radial geodesic with $E=1$, from \eq{p-EOM}, we have 
\begin{equation}
\begin{split}
&\frac{dt}{d\tau}=\frac{1}{f(r)} \\
&\frac{dr}{d\tau}=\pm \sqrt{1-f(r)}
\label{tm1}
\end{split}
\end{equation}
where $\tau$ is a proper time and $\pm$ corresponds to radially outgoing and ingoing 
particles respectively.

\subsubsection{\bf Proper time}
The proper time as measured in the reference frame attached to the particle when 
it travels from $r=r_{\rm i}$ to $r=r_{\rm f}$ can be obtained by integrating the \eq{tm1}
and is given by
\begin{equation}
\begin{split}
\tau(r_{\rm i}\rightarrow r_{\rm f}) &=\pm \int^{r_{\rm f}}_{r_{\rm i}} \frac{1}{\sqrt{1-f(r)}}dr \\
&=\pm \frac{1}{3}\sqrt{\frac{2}{M}} 
\left[\left(r-\frac{Q^2}{2M}\right)^{\frac{1}{2}}\left(r+\frac{Q^2}{M}\right)\right]^{r_{\rm f}}_{r_{\rm i}},
\label{ptime}
\end{split}
\end{equation}
where 
$\pm$ corresponds to the case where $r_{\rm f}>r_{\rm i}$ and $r_{\rm f}<r_{\rm i}$,  i.e., 
when particle moves radially onwards and radially inwards respectively.
\newline

\textbf{Extremal black hole}
\newline

The proper time required for the neutral particle to reach horizon 
from the initial location $r=r_{i}$ using \eq{ptime}
is given by
\begin{equation}
 \tau(r_i\rightarrow M)=\frac{1}{3}\sqrt{\frac{2}{M}} 
\left[\left(r-\frac{M}{2}\right)^{\frac{1}{2}}\left(r+M\right)\right]_M^{r_{\rm i}}
-\frac{2}{3}M
\end{equation}
which is clearly finite. The proper time required for the charged 
particle to reach the horizon however diverges as discussed 
earlier since its effective potential for the radial motion as well as 
its derivative goes to zero at the horizon.
\newline

\textbf{Naked singularity}
\newline

In the naked singularity case, one of the particles starts out as an ingoing particle at 
$r=r_{\rm i}$, gets reflected back at $r=r_{\rm refl}=Q^2/2M$ 
due to the repulsive effect of the naked singularity and arrives at the collision
point $r=r_{\rm min}=Q^2/M$ as an outgoing particle. 
The proper time required in its rest frame from \eq{ptime} is given by 
\begin{equation}
\begin{split}
\tau_1 &= \tau(r_{\rm i} \rightarrow r_{\rm refl})+\tau(r_{\rm refl} \rightarrow r_{\rm min}) \\
&=\frac{2Q^3}{3M^2}+\frac{1}{3}\sqrt{\frac{2}{M}} 
\left[\left(r_{\rm i}-\frac{Q^2}{2M}\right)^{\frac{1}{2}}\left(r_{\rm i}+\frac{Q^2}{M}\right)\right]
\label{pt1}
\end{split}
\end{equation}
The second particle starts out at $r=r_{\rm i}$ and reaches $r_{\rm refl}$ as an ingoing particle
where it collides with the first particle.
The proper time required in its rest frame is given by 
\begin{equation}
\begin{split}
\tau_2 &= \tau(r_{\rm i} \rightarrow r_{\rm min}) \\
&= -\frac{2Q^3}{3M^2}+\frac{1}{3}\sqrt{\frac{2}{M}} 
\left[\left(r_{\rm i}-\frac{Q^2}{2M}\right)^{\frac{1}{2}}\left(r_i+\frac{Q^2}{M}\right)\right]
\label{pt2}
\end{split}
\end{equation}
It is evident from \eq{pt1},\eq{pt2} that the proper time required for the collision is finite in the rest 
frame of both the particles.

\subsubsection{\bf Coordinate time}

We now compute the coordinate time required for the collision 
as measured by the static distant observer in the extremal black hole
and naked singularity cases.
From \eq{tm1} we get 
\begin{equation}
 \frac{dr}{dt}=\pm f(r)\sqrt{1-f(r)}
\end{equation}
The time observed by the distant observer as the particle moves 
from $r=r_{i}$ to $r_{f}$ can be obtained by integrating 
the equation above and is given by 
\begin{equation}
\begin{split}
 T(r_{\rm i} \rightarrow r_{\rm f})&=\pm \int^{r_{\rm f}}_{r_{\rm i}}\frac{1}{f(r)\sqrt{1-f(r)}}dr \\
&=\pm\left[B(r_{\rm f})-B(r_{\rm i})\right]
\label{tb}
\end{split}
\end{equation}
where $\pm$ stands for the radially outgoing and radially ingoing 
particles with $r_{\rm f}>r_{\rm i}$ and $r_{\rm f}<r_{\rm i}$ respectively
as stated earliar, and $B(r)$ is the indefinite integral 
$$B(r)=\int^r\frac{dr}{f(r)\sqrt{1-f(r)}}. $$
\newline
\textbf{Extremal black hole}
\newline

We now compute the coordiate time required for the neutral particle 
to reach the event horizon of the extremal
Reissner-Nordstr\"{o}m black hole. In this case the function $B(r)$ is given by the expression
\begin{equation}
\begin{split}
 B(r)=& M\left( 2 \ln \left|\frac{\sqrt{r-\frac{M}{2}}-\sqrt{\frac{M}{2}}}{\sqrt{r-\frac{M}{2}}+
\sqrt{\frac{M}{2}}}\right|-\frac{\sqrt{2M(r-\frac{M}{2})}}{r-M}\right)\\ 
&+\frac{1}{3}\sqrt{\frac{2}{M}\left(r-\frac{M}{2}\right)} \left(r+7m \right)
\label{tbh}
\end{split}
\end{equation}

Thus it follows from Eqs.~\eq{tb} and \eq{tbh} that the time required 
for the ingoing neutral particle to reach $r_{\rm f}$ diverges 
in the limit $r_{\rm f} \rightarrow M$  as 
\begin{equation}
 T\simeq M \left(\frac{r_{\rm f}-M}{M}\right)^{-1}
\label{tbh1}
\end{equation}

The center of mass energy of collision $E_{\rm cm}$ 
between the charged and uncharged particles as a function of the
collision location $r_{\rm f}$ varies as \cite{Nakao}
\begin{equation}
E_{\rm cm} \simeq \sqrt{2}m \left(\frac{r_{\rm f}-M}{M}\right)^{-\frac{1}{2}}  
\label{ecmbh}            
\end{equation}
where $m$ is the mass of each of the colliding particles.

It follows from Eqs.~\eq{tbh1} and \eq{ecmbh} that the time required for the neutral particle 
to participate in the collision at the radial location $r=r_{\rm f}\rightarrow M$ is thus given by 
\begin{equation}
\begin{split}
 T & \simeq \frac{M}{2} \left(\frac{E_{\rm cm}}{m}\right)^{2} \\
& \simeq 1.3\times10^{25} \left(\frac{M}{M_{\rm \odot}}\right)\left(\frac{E_{\rm cm}}{E_{\rm pl}}\right)^{2}
\left(\frac{m_{\rm p}}{m}\right)^2 \ {\rm yr},
\end{split}
\end{equation}
where $M_{\rm \odot}$ is the solar mass, $E_{\rm pl}$ is Planck energy and $m_{\rm p}$ is mass of the proton.
The time required for the neutral particle with mass $m_{p}$ such as neutron, to participate 
in a Planck scale collision around a solar mass extremal black hole is 
approximately $10^{15}$ times larger than 
the age of the universe. The time required for the charged particle 
to reach the collision point will be even larger. Therefore the phenomenon of ultra-high energy 
collisions around charged black holes does not occur within the Hubble time 
scale and thus has no observable consequences whatsoever.
\newline

\textbf{Naked singularity}
\newline

We now discuss the timescale associated with 
the ultra-high energy collision around the Reissner-Nordstr\"{o}m
naked singularity. The function $B(r)$ in this case is given by the following expression.
\begin{equation}
\begin{split}
 B(r) &=M\ln\left| \frac{r-\sqrt{2rM-Q^2}}{r+\sqrt{2rM-Q^2}}\right|\\
&+\frac{\left(2M^2-Q^2\right)}{\sqrt{Q^2-M^2}}\arctan\left(\frac{\sqrt{2rM-Q^2}-M}{\sqrt{Q^2-M^2}}\right)\\
&+\frac{\left(2M^2-Q^2\right)}{\sqrt{Q^2-M^2}}\arctan\left(\frac{\sqrt{2rM-Q^2}+M}{\sqrt{Q^2-M^2}}\right)\\
&+\frac{\sqrt{2rM-Q^2}}{3M^2}\left(rM+Q^2+6M^2\right) 
\label{nsb}
\end{split}
\end{equation}


The time required for the ingoing neutral particle starting at $r=r_{\rm i}$ 
to get reflected at $r=r_{\rm refl}=Q^2/2M$
as an outgoing particle and to reach the collision point $r=r_{\rm min}=Q^2/M$ 
in the limit $Q\rightarrow M$, from Eqs.~\eq{tb} and \eq{nsb} is given by 
\begin{equation}
\begin{split}
 T_1 &= T(r_{\rm i}\rightarrow r_{\rm refl})+T(r_{\rm refl}\rightarrow r_{\rm min}) \\
&\simeq \frac{3\pi}{2}\frac{M^2}{\sqrt{Q^2-M^2}}
\label{t1}
\end{split}
\end{equation}
Whereas the time required for the second ingoing neutral particle 
to reach $r=r_{\rm min}$ starting from $r=r_{\rm i}$, from Eqs.~\eq{tb} and \eq{nsb} is given by
\begin{equation}
\begin{split}
 T_2 &= T(r_{\rm i}\rightarrow r_{\rm min}) \simeq \frac{\pi}{2}\frac{M^2}{\sqrt{Q^2-M^2}}
\label{t2}
\end{split}
\end{equation}
It is clear from \eq{t1},\eq{t2} that $T_1$ and $T_2$ 
diverge as $1/\sqrt{Q^2-M^2}$ in the limit $Q \rightarrow M$.

The center of mass energy of collision between two particles at $r=r_{\rm min}=Q^2/M$ 
in the Reissner-Nordstr\"{o}m naked singularity case 
in the limit $Q \rightarrow M$ is given by Eq.~\eq{ens}
\begin{equation}
 E_{\rm cm}\simeq \frac{2mM}{\sqrt{Q^2-M^2}}
\label{ecmns}
\end{equation}
where $m$ is the mass of each of the colliding particles.

From Eqs.~\eq{t1}, \eq{t2} and \eq{ecmns}, the time scale associated with the collision is given by 
\begin{equation}
\begin{split}
T &\simeq T_{2} \simeq \frac{T_{1}}{3} \simeq \frac{\pi}{4} M \left(\frac{E_{\rm cm}}{m}\right) \\
&\simeq 2.32 \times 10^{6} \left(\frac{M}{M_{\rm \odot}}\right)\left(\frac{E_{\rm cm}}{E_{\rm pl}}\right)
\left(\frac{m_{\rm p}}{m}\right) \ yr
\end{split}
\end{equation}
where as before $M_{\rm \odot}$ is mass of the sun, $E_{\rm pl}$ is 
the Planck energy and $m_{\rm p}$ is mass of the proton.
We see that the time scale associated with the Planck scale collision of two
neutrons 
around a solar mass naked singularity
is merely of the order of million years which is $10^{4}$ times smaller than 
the age of the universe.

This implies that the trans-Plankian collisions around naked 
singularities are conceivable and might be
observable either in our galaxy or
at very high cosmological redshifts.
Furthermore if the particles
continuously accrete from a distant location $r=r_{\rm i}>r_{\rm min}$, in a steady state,
the rate of occurrence of the collisions will be same as 
the accretion rate. Thus one could say that there is no inflatiing 
time-scale problem in the naked singular Reissner-Nordstr\"{o}m
spacetime while it does exists in the extremal black hole geometry.

\section{Acceleration of shells by Reissner-Nordstr\"{o}m naked singular geometry}
 In this section, we discuss the validity of test particle approximation
on the particle collision around Reissner-Nordstr\"{o}m naked singular geometry.
We should consider two type of ``back reaction'', i.e., the  
effects of the gravitational radiation and the conservative self-force.

As long as we consider the radially moving particles, the effect of gravitational radiation 
does not change it to non-radial one. However, if the energy of the particle 
is released by the gravitational radiation, initially marginally bound particle will be 
bound.  We denote the energies of the particles by $E_1$ and $E_2$. If the 
gravitational emission is negligible, $E_1$ and $E_2$ are constants of motion, but 
it might vary with time if the gravitational emission is  not negligible. 
The 4-velocities of the particles are written in the form
\begin{eqnarray}
U_1^\mu&=&\left(\frac{E_1}{f},\sqrt{E_1^2-f},0,0\right), \\
U_2^\mu&=&\left(\frac{E_2}{f},-\sqrt{E_2^2-f},0,0\right).
\end{eqnarray}
As before, we assume that the collision occurs at the minimum of $f$, i.e., 
$f=\epsilon(2-\epsilon)$, 
where $\epsilon$ has been defined by Eq.~\eq{e-def}, 
and then the collision energy at the center of mass frame is given by
\begin{equation}
E_{\rm cm}^2=\frac{2m^2}{f}\left[f+E_1E_2+\sqrt{(E_1^2-f)(E_2^2-f)}
\right]
\end{equation} 
Since the $r$-components of $U_1^\mu$ and $U_2^\mu$ should be real, $E_1$ and $E_2$ 
should be larger than or equal to $\sqrt{\epsilon(2-\epsilon)}$. If  
$E_1$ and $E_2$ become several times $\sqrt{\epsilon(2-\epsilon)}$ 
by the emission of the gravitational radiation, the collision 
energy $E_{\rm cm}$ takes small value which is several times $m$. 
However note that in this case the emitted gravitational radiation would be so large that the conserved
energies, which were assumed to have unit value to begin with, drastially reduce to a value that 
is nearly equal to zero.
It is beyond the scope of this paper to estimate how large is the
amount of energies of the particles are released by the gravitational radiation, 
and hence, we cannot make any quantitative statement. 
If the colliding particles do not drastically loose the energies to a value close to zero
and carry a descent fraction of the initial energies, the ultra-high energy collision can occur. 

If so, it is sufficient to consider the effect of conservative self-force.
Consideration of conservative self-force is important since it 
can turn a near extremal naked singular configuration into  a black hole and thus
hiding the ultra-high energy collisions below the event horizon.

In this section, 
since to treat conservative self-force for the point particle is difficult, 
we study analogous system, i.e., 
the collision of  spherical shells 
in the Reissner-Nordstr\"{o}m naked singular geometry.
It is also well justified on the physical grounds, since in a realistic situation the 
accretion of the matter onto 
a massive compact object would be more or less isotropic. Therefore the amount of gravitational radiation 
emitted per particle will be significantly reduced \cite{Nakamura} and its effect on the process of 
ultra-high energy collisions can be ignored to a very good approximation. Thus would suffice 
to consider only the conservative self-force.
The dynamics of the spherical thin shells is 
tractable exactly owing to the spherical symmetry of the system. 
Due to the gravity generated by the shells themselves, 
the equations describing the motion of shells are no longer 
the geodesic equations in the Reissner-Nordstr\"{o}m spacetime.

We deal with the situation that is analogous to the scenario 
described in the previous section, in order to draw a parallel to 
and compare with the test particle case.  
We assume that the deviation of the charge from the mass 
associated with the naked singularity is vanishingly small.

\subsection{Junction conditions}

We first describe the procedure to deal with the thin shells with taking 
into account their gravity \cite{Israel}.
We basically follow notation and convention of Ref.~\cite{Poisson}. 
A trajectory of a shell that is being considered here is a timelike hypersurface 
with a thin surface layer of matter in the four dimensional ambient 
spacetime manifold: we denote it by $\Sigma$. Then,  
a shell $S$ means an intersection between $\Sigma$ 
and a spacelike hypersurface with 
constant time coordinate chosen appropriately.  
Since the finite amount of energy  
exists within the infinitesimally thin layer, 
the energy-momentum tensor is infinite, 
but it is possible to define it as a distribution. 

The geometry of a hypersurface can be described 
by specifying a three dimensional 
metric $h_{ab}$ within it (also known as the induced metric) 
and an extrinsic curvature $K_{ab}$, which is a three dimensional tensor 
describing how the  hypersurface is embedded in the ambient spacetime. 
Even if the trajectory of the shell is a singular hypersurface, we assume 
that the metric of four dimensional spacetime is everywhere continuous. 
Thus, the induced metric $h_{ab}$ of the shell 
is assumed to be continuous. By contrast, the extrinsic curvature 
$K_{ab}$ of the shell may be discontinuous 
across the shell due to the distributional energy-momentum 
tensor on the shell. 
$\Sigma$ separates the spacetime in two regions 
${\cal V}_{1}$ and ${\cal V}_{2}$. 
The coordinates defined in these two regions is denoted by 
$x^{\mu}_J$ ($J=1,2$), whereas the coordinates within $\Sigma$ 
is denoted by $y^{a}$  ($a=0,1,2)$. 
Although the metric is continuous, the components of it may not be continuous, 
since the coordinate systems may be discontinuous at $\Sigma$. 

The projection operator 
from the four dimensional ambient spacetime to $\Sigma$ is given by
\begin{equation}
e^{\mu}_{Ja}=\frac{\partial x_{J}^{\mu}}{\partial y^{a}}.
\end{equation}
Here, the index $J$ of the projection operator indicates side of $\Sigma$ on which 
the quantity is defined, ${\cal V}_{1}$ or ${\cal V}_{2}$. 
Denoting the components of the metric by $g_{J\mu\nu}$,  
the induced metric on the hypersurface is given by
\begin{equation}
h_{ab}=g_{J\mu\nu} e^{\mu}_{Ja}e^{\nu}_{Jb}.
\end{equation}
The extrinsic curvature of the shell is given by
\begin{equation}
K_{Jab}=e^{\mu}_{Ja}e^{\nu}_{Jb}n_{J\mu;\nu} \label{K-def}
\end{equation}
where $n_{J\mu;\nu}$ denotes a component of the 
covariant derivative of the unit normal vector $n_{J\mu}$ to $\Sigma$, which 
is directed from ${\cal V}_1$ to ${\cal V}_2$.  Here, note that the unit normal vector 
to $\Sigma$ is unique, since the metric tensor is everywhere continuous. 

By denoting the components of the energy momentum tensor 
of the shell by $T_J^{\mu\nu}$, it is given by the following form
\begin{equation}
T_J^{\mu\nu}=\delta(\lambda)S^{ab}e_{Ja}^{\mu}e_{Jb}^{\nu}, \label{T1}
\end{equation}
where $S^{ab}$ is a three dimensional tensor defined over $\Sigma$ 
shell, which is called the surface energy-momentum tensor,  
$\delta(\lambda)$ is Dirac delta function, 
and $\lambda$ is the Gaussian normal coordinate which is equal to zero 
on $\Sigma$.

The junction condition is given in the form of the condition on the discontinuity of 
the extrinsic curvature of $\Sigma$ as follows; 
\begin{equation}
K_{2ab}-K_{1ab}=-8\pi \left(S_{ab}-\frac{1}{2}h_{ab}S^c_c\right). \label{EOM}
\end{equation}

\subsection{Motion of a neutral dust shell}

\begin{figure}
\begin{center}
\includegraphics[width=0.3\textwidth]{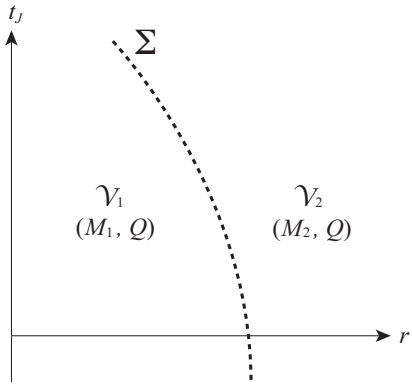}
\caption{\label{1-shell}
This is schematic diagram of the spherically symmetric spacetime divided into 
two domains ${\cal V}_1$, ${\cal V}_2$ by the trajectory of 
a thin shell $\Sigma$ depicted by a dashed curve. 
The spacetime metric in the two domains ${\cal V}_1$, ${\cal V}_2$ 
is Reissner-Nordstr\"{o}m with different values of mass parameters, 
namely $M_1$ and $M_2$, but with the same charge $Q$.
}
\end{center}
\end{figure}

We now consider a case where the both regions ${\cal V}_1$ and ${\cal V}_2$ 
are the Reissner-Nordstr\"{o}m and the shell $S$ is spherically symmetric and 
made up of charge neutral dust. 
Due to the charge neutrality of the shell, the charge parameters in the both regions 
are identical, and we denote it by $Q$. 
By contrast, the value of the mass parameter would be different 
in two regions (see Fig.~\ref{1-shell}). 

We use the coordinate systems $x_{1}^{\mu}=(t_1,r,\theta,\phi)$ and 
$x_{2}^{\mu}=(t_2,r,\theta,\phi)$ in the region ${\cal V}_1$ and ${\cal V}_2$, 
respectively. Note that the time coordinate is not continuous, 
whereas $r$, $\theta$ and $\phi$ are everywhere continuous. 
The metric in these two regions can be written as
\begin{equation}
ds^2=-f_J(r) dt_J^2+\frac{1}{f_J(r) }dr^2+r^2(d\theta^2+\sin^2\theta d\phi^2),
\end{equation}
where
\begin{equation}
f_J(x)=1-\frac{2M_J}{x}+\frac{Q^2}{x^2},  \label{f1-def}.
\end{equation}
The Misner-Sharp mass of the shell is defined by
\begin{equation}
\mu:=M_2-M_1,
\end{equation}
and we assume $\mu>0$. The positivity of $\mu$ naturally introduce a picture 
that ${\cal V}_1$ is the inside of the shell $S$, whereas ${\cal V}_2$ is the outside. 

We use the coordinates $(\tau,\theta,\phi)$ on $\Sigma$, where 
$\tau$ is taken to be the proper time for an observer comoving with the shell. 
The intrinsic metric of the shell is then written as 
\begin{equation}
ds_{\Sigma}^2=-d\tau^2+R(\tau)^2\left(d\theta^2+\sin^2\theta d\phi^2\right) \label{IM1}
\end{equation}
The proper time of the shell can parametrize the trajectory of the shell, i.e., 
\begin{equation}
t_J=T_J(\tau)~~~~{\rm and}~~~~r=R(\tau).
\end{equation}
The projection operator is then given by
\begin{eqnarray}
e^\mu_{J\tau}&=&\left(\dot{T}_J,\dot{R},0,0\right), \\
e^\mu_{J\theta}&=&\left(0,0,1,0\right), \\
e^\mu_{J\phi}&=&\left(0,0,0,1\right).
\end{eqnarray}
The induced metric is given by
\begin{eqnarray}
ds_{\Sigma}^2&=&g_{\mu\nu}e^\mu_{J a}e^\nu_{Jb}dy^ady^b \cr
&=&\left[-f_J(R)\dot{T_J}^2+\frac{1}{f_J(R)}\dot{R}^2\right]d\tau^2 \cr
&& \cr
&+&R(\tau)^2
(d\theta^2+\sin^2\theta d\phi^2), \label{IM2}
\end{eqnarray}
where a dot represents a derivative with respect to $\tau$. 
Equations \eq{IM1} and \eq{IM2} imply 
\begin{equation}
f_J(R)\dot{T_J}=\sqrt{\dot{R}^2+f_J(R)}=:\beta_J(R,\dot{R})
\label{tnsame}
\end{equation}
Equation \eq{tnsame} implies that the time coordinate 
in the regions ${\cal V}_1$ and ${\cal V}_2$ necessarily have to be different.

The unit normal vector $n_{J\mu}$ to $\Sigma$ is given by
\begin{eqnarray}
n_{J\mu}=\left(-\dot{R}, \dot{T}_J,0,0\right) =\left(-\dot{R},\frac{\beta_J}{f_J},0,0\right)
\end{eqnarray}
Substituting the above expression into Eq.~\eq{K-def}, we find that 
the non-vanishing components of the extrinsic curvature are given by
\begin{equation}
K^{\tau}_{J\tau}=\frac{\dot\beta_J}{\dot{R}}, ~~~~
K^{\theta}_{J\theta}=K^{\phi}_{J\phi}=\frac{\beta_J}{R}.  \label{Kab}
\end{equation}

The energy-momentum tensor of the dust within the thin shell is given by
\begin{equation}
T_J^{\mu\nu}=\sigma(\tau) \delta(\lambda)U_J^{\mu}U_J^{\nu}
\end{equation}
where $\sigma$ is the surface density and $U_J^{\mu}$ is a component of 
the 4-velocity of the dust, 
which is equivalent to $e_{J\tau}^\mu$. We assume that $\sigma$ is non-negative. 
Comparing the above equation with Eq.~\eq{T1}, we get
\begin{equation}
S^{ab}=\sigma u^{a}u^{b}, \label{Sab}
\end{equation}
where $u^a$ is the 3-velocity of the dust within the shell, whose components 
are given by 
\begin{equation}
u^a=(1,0,0).
\end{equation}
By using Eqs. \eq{Kab} and \eq{Sab}, we now write down Eq.~\eq{EOM} and obtain
\begin{eqnarray}
-\sigma=\frac{\beta_2-\beta_1}{4\pi R} \label{meq}\\
0=\frac{\beta_2-\beta_1}{R}+\frac{\dot{\beta_2}-\dot{\beta_1}}{\dot{R}}\label{vel}
\end{eqnarray}

Equations \eq{meq} and \eq{vel} taken together give
\begin{equation}
4\pi R^2 \sigma=m={\rm constant}. 
\end{equation}
The constant $m$ is interpreted as the proper mass of the shell, 
and we get an energy equation for the shell as follows
\begin{eqnarray}
\dot{R}^2&=&\frac{1}{m^2}\left(\mu+\frac{m^2}{2R}\right)^2-f_{1}(R) \cr
&=&\frac{1}{m^2}\left(\mu-\frac{m^2}{2R}\right)^2-f_{2}(R) \label{EOM2}
\end{eqnarray}
As in the case of the test particle, let us introduce the following effective potential 
\begin{equation}
V_{\rm eff}=1-\frac{2\langle M\rangle}{R}+\frac{Q^2}{R^2}-\left(\frac{m}{2R}\right)^2,
\end{equation}
where 
\begin{equation}
\langle M\rangle=\frac{M_1+M_2}{2}.
\end{equation}
Then, Eq.~\eq{EOM2} is written in the very similar form to Eq.~\eq{p-EOM} 
for the test particle as
\begin{equation}
\left(\frac{dR}{d\tau}\right)^2+V_{\rm eff}=E^2,
\end{equation}
where $E=\mu/m$ is the energy of the shell per unit proper mass. 

\begin{figure}[htbp]
\begin{center}
\includegraphics[width=0.43\textwidth]{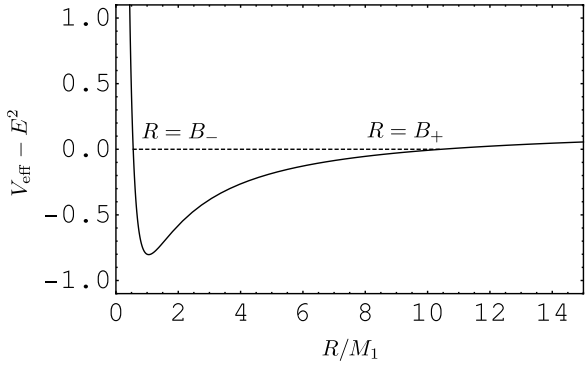}
\caption{The effective potential $V_{\rm eff}$ minus the square of 
specific energy $E^2$ of the shell is depicted for the case 
of $m<2Q$ and $E<1$.  The allowed domain for the motion of the shell is specified by the dashed line. 
}
\label{B_ml2Q}
\end{center}
\end{figure}

\begin{figure}[htbp]
\begin{center}
\includegraphics[width=0.43\textwidth]{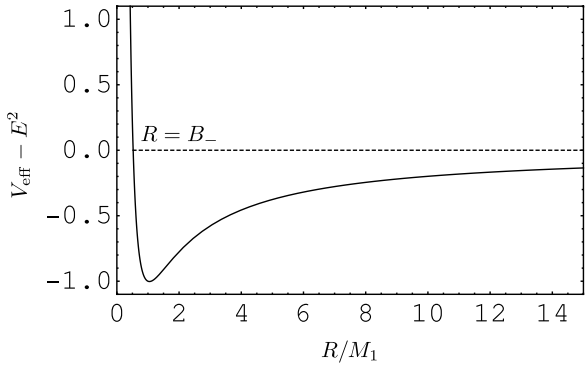}
\caption{The same as Fig.~\ref{B_ml2Q}, but $E>1$.  
}
\label{unB_ml2Q}
\end{center}
\end{figure}

\begin{figure}[htbp]
\begin{center}
\includegraphics[width=0.4\textwidth]{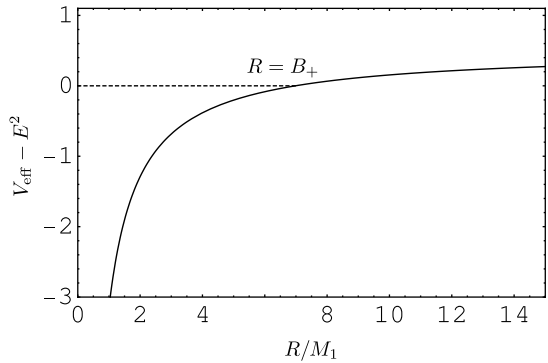}
\caption{The same as Fig.~\ref{B_ml2Q}, but $m>2Q$.  
}
\label{B_mg2Q}
\end{center}
\end{figure}

\begin{figure}[htbp]
\begin{center}
\includegraphics[width=0.4\textwidth]{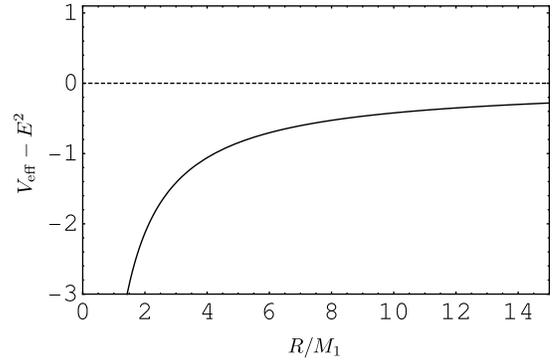}
\caption{The same as Fig.~\ref{B_ml2Q}, but $m>2Q$ and $E>1$.  }
\label{unB_mg2Q}
\end{center}
\end{figure}

We depict the effective potential $V_{\rm eff}$ for the shell as a function of $R$ in Figs.~3-6.  
First, we consider the case of  $m<2Q$. For $E<1$ (see Fig.~\ref{B_ml2Q}), the motion of the shell 
is restricted within the domain $B_-\leq R \leq B_+$, where  
\begin{equation}
B_\pm=\frac{\langle M\rangle}{1-E^2}\left(1\pm\sqrt{1-\frac{Q^2-m^2/4}{\langle M\rangle^2}
\left(1-E^2\right)}\right).
\label{B-def}
\end{equation}
In the case of $E=\sqrt{1-\langle M\rangle^2/(Q^2-m^2/4)}$, $B_+$ is equal to $B_-$, and 
the particle stays stably at rest at the radius $R=(Q^2-m^2/4)/\langle M\rangle$. 
For $E\geq1$ (see Fig.~\ref{unB_ml2Q}), the allowed domain for the
motion 
is $R\geq B_-$. Initially outgoing shell 
monotonically  approaches to infinity, $R\rightarrow\infty$, whereas initially ingoing 
shell turns to be outgoing. These behaviors are basically the same as that of the test particle. 
The repulsive nature of the charged singularity halts the collapse of the shell. 

By contrast, in the case of $m\geq2Q$, the allowed domain for the motion  
is $R \leq B_+$ for $E<1$ (see Fig.~\ref{B_mg2Q}) and thus the shell with $E<1$ 
necessarily collapses to the singularity at $r=0$. 
In the case of $E\geq1$ (see Fig.~\ref{unB_mg2Q}), whole domain 
is allowed for the motion of the shell; the initially outgoing shell goes to infinity, whereas 
the initially ingoing shell collapses to the singularity at $r=0$. The
self-gravity of the shell overcomes the
repulsive gravity of the charged singularity. 

We rewrite the effective potential in the form
\begin{equation}
V_{\rm eff}(r)=f_2(r)+E^2-\left(E-\frac{m}{2r}\right)^2.
\end{equation}
From the above equation, we have
\begin{equation}
\dot{R}^2=-V_{\rm eff}(R_{+})+E^2=\left(E-\frac{m}{2R_{+}}\right)^2\geq0,
\end{equation}
where $R_{+}$ is a larger root of $f_2(R_+)=0$. Thus, in the case of 
$M_2 \geq Q$, an ingoing shell necessarily enters into the black hole 
and goes to the another asymptotically flat region. By contrast, in the case 
of $M_2<Q$, there is only one asymptotically flat region, and hence 
even in the case of the ingoing shell, the shell remains in this asymptotically flat reg,ion 
as long as it does not hit the spacetime singularity at $r=0$. The situation is similar to 
the case of a radially moving test particle.

\subsection{Collision of charge neutral shells}

\begin{figure}
\begin{center}
\includegraphics[width=0.4\textwidth]{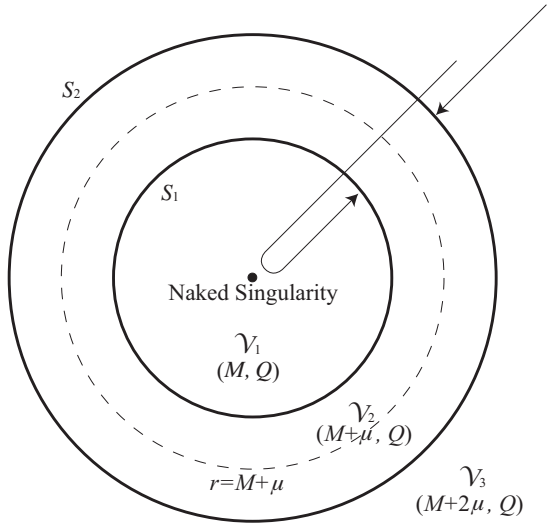}
\caption{\label{2-shell}
This is a schematic diagram showing the motion and collision of two shells 
$S_1$ and $S_2$. 
There is a Reissner-Nordstr\"{o}m naked singularity  
at the center whose charge is slightly larger than the mass. The shell $S_1$ which 
is initially ingoing turns back as an outgoing shell and then collides with the ingoing shell 
$S_2$ at $r=M+\mu$. The similar picture also can be drawn in case of particle collision 
replacing shells by particles.
}
\end{center}
\end{figure}

Now we describe the process of acceleration and collision of charge neutral shells 
whose motion has been considered in the preceding section. 
We consider two concentric spherical thin shells $S_1$ and $S_2$. 
These  shells divide the spacetime into three regions 
each of which is denoted by ${\cal V}_J$ $(J=1,2,3)$: 
$S_1$ faces ${\cal V}_1$ and ${\cal V}_2$, whereas $S_2$ faces ${\cal V}_2$ and ${\cal V}_3$ 
(see Fig.~\ref{2-shell}). 
The metric in the three regions would be given by Reissner-Nordstr\"{o}m geometry 
with the different values of parameters in three regions. 
The shells are assumed to be electrically neutral and thus the charge parameters 
in the three regions take an identical value $Q$. 
By contrast, mass parameters take different values in three regions. We denote them by 
$M_J$. For simplicity, we assume that these shells have identical Misner-Sharp mass $\mu$, 
and they are given by $M_1=M$, $M_2=M+\mu$ and $M_3=M+2\mu$ with $\mu>0$.
Further, we assume that the charge $Q$ is somewhat larger than $M_3$. 
Here, we replace the define a small parameter $\epsilon$ by 
\begin{equation}
Q-M=:\epsilon Q. \label{e-def-2}
\end{equation}
By the assumption, we have $0<\epsilon \ll 1$, and by another assumption $M_3<Q$, 
we have $\mu<\epsilon Q/2$. Thus, we may write $\mu$ in the form
\begin{equation}
\mu =\frac{\epsilon Q}{2}\hat{\mu}~~{\rm with}~~0<\hat{\mu}< 1.
\label{mu-def}
\end{equation}
The above condition ensures that the naked singularity 
does not turn into a black hole by the shells $S_1$ and $S_2$.

Hereafter, for simplicity, we 
assume that the shells are marginally bound, i.e., $\mu=m$.
Following exactly the same procedure  as in the one-shell case, 
the radial components of 4-velocities of $S_J$ ($J=1,2$) whose radii 
are denoted by $R_J$ can now be written as
\begin{eqnarray}
\dot{R_J}&=&\pm\sqrt{\left(1+\frac{m}{2R_J}\right)^2-f_J(R_J)} \cr
&=&\pm\sqrt{\left(1-\frac{m}{2R_J}\right)^2-f_{J+1}(R_J)}
\end{eqnarray}
where  $f_J$ are identical to Eq. (\ref{f1-def}), and 
$\pm$ stands for outgoing and ingoing shell, respectively.
Using the normalization condition for 4-velocity $\bm{U}_J\cdot\bm{U}_J=-1$, 
we obtain the time components of the 4-velocity with respect to the coordinate 
basis in the domain ${\cal V}_2$, 
\begin{equation}
\nonumber
\dot{T_J}=\frac{1}{f_2(R_J)}\sqrt{\dot{R_J}^2+f_2(R_J)}. 
\end{equation}

The radial components of 4-velocities of both $S_1$ and $S_2$ 
would go to zero at infinity by the assumption of $\mu=m$. 
By carefully taking a limit $E\rightarrow1$ for $B_-$ in Eq.~\eq{B-def}, 
we have the turning points $R=B_{1-}$ for $S_1$ and $R=B_{2-}$ for $S_2$, where 
\begin{eqnarray}
B_{1-}&=&\frac{Q^2-m^2/4}{2M+m}\approx \frac{Q^2}{2M}\approx\frac{M}{2}, \\
B_{2-}&=&\frac{Q^2-m^2/4}{2(M+m)}\approx \frac{Q^2}{2M}\approx\frac{M}{2}
\end{eqnarray}
Both the turning points are the same order, but $B_{1-}>B_{2-}$.

We consider a situation where the inner shell $S_1$ starts off at infinity as an ingoing shell; $S_1$ 
then turns back at $R\approx M/2$ and emerges as an outgoing shell and 
collides with the outer ingoing shell $S_2$ at $R \approx Q^2/M \approx M$. 
This situation is exactly analogous to the situation encountered in the case of 
test particles in Sec.~II. 

The energy of two shells at ``the center of mass frame" was defined 
in \cite{Nakao} in a following way by generalizing the definition of the center of mass 
energy of the particles. In case of the particle collisions, 
in order to compute the center of mass energy, 
one goes to the orthonormal frame in which the spatial components of the total momentum 
of the two particles is zero. The time component yields the center of mass energy.
While dealing with the collision event of the shells, the center of mass frame was defined 
to be an orthonormal frame in which the energy flux along 
the spatial direction is zero and the center of mass energy 
is defined analogously. We obtain for the shells 
\begin{equation}
E_{\rm cm}^2=2m^2\left(1-\bm{U}_1\cdot \bm{U}_2 \right)
\end{equation}

We can compute $\bm{U}_1\cdot \bm{U}_2$ by using their components 
with respect to the coordinate basis in the region ${\cal V}_2$, 
and we have the center of mass energy 
of collision at any given value of $R$ as 
\begin{widetext}
\begin{eqnarray}
E_{\rm cm}^2=2m^2
\left[1- \frac{1}{f_2}\left(\dot{R}_1\dot{R}_2-\sqrt{(\dot{R}^2_1+f_2)(\dot{R}_2^2+f_2)}\right)\right]
\label{Ecm}
\end{eqnarray}
\end{widetext}
The circumferential radius at the the minimum of $f_2$ is $R=Q^2/M_2$, 
and let us consider the collision there. 
From Eqs.~(\ref{e-def-2}) and (\ref{mu-def}), we have 
\begin{equation}
Q-M_2=\epsilon Q\left(1-\frac{\hat{\mu}}{2}\right).
\end{equation}
Then, If the signs of $\dot{R}_1$ and $\dot{R}_2$ are  different from each other, 
we have, for $0<\epsilon\ll1$, 
\begin{equation}
E_{\rm cm}^2|_{R=Q^2/M_2}\simeq \frac{4m^2}{(2-\hat{\mu})\epsilon}.
\label{Ecm-max}
\end{equation}
We can see from the above equation that as in the case of the test particles, 
the energy of two spherical shells 
at the center of mass frame can be arbitrarily large.   

Here, we assume that a shell is composed of $N$ particles each of which has a mass
$\delta m=m/N$. The center of mass energy $E_{\rm p}$ of a collision
between two of  constituent particles is given by
\begin{equation}
E_{\rm p}^2=\frac{\delta m^2}{m^2}E_{\rm cm}^2.
\end{equation}
Using the above equation and Eq.~(\ref{Ecm-max}), the 
collision energy at $R=Q^2/M_2$ with $0<\epsilon \ll1$ 
is given by
\begin{equation}
E_{\rm p}^2\simeq \frac{4\delta m^2}{(2-\hat{\mu})\epsilon}.
\end{equation}
The above equation seems to imply that the center of mass energy can be indefinitely large.
However, in order that the description by a spherical shell is valid,
the number of particles $N$ should be much larger than unity, i. e.,
\begin{equation}
N=\frac{m}{\delta m}=\frac{Q\hat{\mu}}{2\delta m}\epsilon 
=\frac{M\hat{\mu}}{2\delta m}\frac{\epsilon}{1-\epsilon}\gg1,
\end{equation}
or, by assuming $\delta m/M\ll1$, 
\begin{equation}
\epsilon \gg \frac{2 \delta m}{M\hat{\mu}}.
\end{equation}
Due to this constraint,  we have
\begin{eqnarray}
E_{\rm p}&\ll& \sqrt{\frac{2\hat{\mu}}{2-\hat{\mu}}\delta mM}<\sqrt{2\delta mM} \nonumber \\
&=&4.58\times10^{28}\left(\frac{\delta m}{m_{\rm p}}\right)^{\frac{1}{2}}\left(\frac{M}{M_{\odot}}\right)^{\frac{1}{2}}
{\rm GeV}.
\end{eqnarray}
The above equation implies that if $M$ is order of the solar mass
$M_\odot =1.99\times 10^{30}$kg,
the collision energy $E_{\rm p}$ between particles at the center of mass frame 
can exceed Planck scale $m_{\rm pl}=\sqrt{hc/2G}=2.16\times10^{19}{\rm GeV}$ even if
$\delta m$  is the order of the proton mass $m_{\rm p}=0.938$GeV.

\section{Conclusions}

In this paper, we studied the particle and shell acceleration by Reissner-Nordstr\"{o}m
naked singularities. The phenomenon of particle acceleration and collision with extremely
large energy at the center of mass frame 
was previously studied and explored in the background
of extremal and near extremal black holes.
We extended this result to the near extremal naked singularities.
We showed that there are significant qualitative differences in the particle acceleration
mechanism between black holes and naked singularities.
In case of black, the particle collision between ingoing particles should be considered, and 
in order to achieve large collision energy at the center of mass frame, 
fine tuning of parameters is necessary,
and further the proper time of one of two particles 
required for such a collision is very long. 
On the contrary, in case of naked singularity,
it is possible to consider a collision between ingoing and outgoing particles,
since due to the absence of the event horizon and the repulsive gravity effects near singularity,
initially ingoing particle turns back as an outgoing particle. 
This fact eliminates the necessity of the fine tuning of some parameters 
and also the required proper time required for such a collision
need not be so long.

We also calculate the coordinate time as seen by the observer at
infinity required for the
ultra-high energy collisions to occur for extremal black hole as well
as naked singular geometry. We show that the time
required for the Planck-scale collisions around naked singularity is
of the order of million years which is much smaller 
than the age of the universe. Whereas the time scale in extremal black
hole case in the analogous process is many orders of magnitude
larger than the Hubble time. Therefore the high energy collisions
occuring around the naked singularities, subject 
to their existence will be observable. Rate of occurence of the
collisions will be same as the rate of the accretion of the matter 
in a steady state. On the contrary, in the black hole case high energy 
collisions would not occur within the Hubble time and thus 
would have no observational consequences.

Particles participating in the collision are assumed
to be test particles following the geodesics on the background geometry.
The effects of gravity generated by the particles are ignored.
Thus, to study whether or not the phenomenon of divergence of center of 
mass energy survives, 
we studied the collision between the concentric spherical shells. 
The gravity of the shells is taken into account in an exact calculation, and the 
energy of collision between shells at ``the center of mass frame" 
is computed in a situation analogous to the test particle case.
It is shown that, in this case, due to the condition that the outermost region is described
by the over-charged RN spacetime, the center of mass energy of a collision
between two of the constituent particles of the shells is bounded above. However, if the mass
of the central naked singularity is order of the solar mass,
and if the mass of a 
constituent particle of the shells is order of the proton mass,
the upper bound exceeds $10^{28}$GeV which is much larger 
than the Planck scale.

\section*{Acknowledgments}
MK is supported by the JSPS Grant-in-Aid for Scientific Research No.23$\cdot$2182. We would like to thank
the anonymous referee for valuable comments that led to an improvement of the manuscript.

\end{document}